\documentclass[prd,nofootinbib,showpacs,preprint,aps]{revtex4-1}
\usepackage{subfigure}
\usepackage{amsmath}
\usepackage{amsfonts}
\usepackage{amssymb}

\def\beq{\begin{equation}}
\def\eeq{\end{equation}}
\def\bsp{\begin{split}}
\def\esp{\end{split}}
\def\bea{\begin{eqnarray}}
\def\eea{\end{eqnarray}}
\def\ba{\begin{array}}
\def\ea{\end{array}}

\def\lb{\left(}
\def\rb{\right)}

\def\l.{\left.}
\def\r.{\right.}

\def\ie{\it ie}

\def\part{\partial}

\def\R{\ensuremath{\mathbb{R}}}

\usepackage[applemac]{inputenc}
\usepackage[T1]{fontenc}
\usepackage{amssymb}
\usepackage{amsfonts}
\usepackage{amsfonts} 
\usepackage{graphicx} 

\usepackage{bbm}
\def\beq{\begin{equation}}
\def\eeq{\end{equation}}
\def\bsp{\begin{split}}
\def\esp{\end{split}}
\def\bea{\begin{eqnarray}}
\def\eea{\end{eqnarray}}
\def\ba{\begin{array}}
\def\ea{\end{array}}

\def\lb{\left(}
\def\rb{\right)}

\def\l.{\left.}
\def\r.{\right.}

\def\part{\partial}

\def\R{\ensuremath{\mathbb{R}}}

\begin{document}
\preprint{UdeM-GPP-TH-20-xxx}
\title{You can see a clock running backwards in time}
\author{M. B. Paranjape} 
\email{paranj@lps.umontreal.ca}
\affiliation{Groupe de physique des particules, D\'epartement de physique, and Centre de recherche mathématiques,
Universit\'e de Montr\'eal,
C.P. 6128, succ. centre-ville, Montr\'eal, 
Qu\'ebec, Canada, H3C 3J7 }
\begin{abstract}\baselineskip=18pt
The epitome of acausal or anti-chronological behaviour would be to see a clock running backwards in time.  In this essay we point out that this is indeed possible, but there is no problem with causality.  What you see isn't what is really happening.  Locally, causality is always respected.  However our observation should be cause for pause to astronomers and cosmologists, who strictly observe events occurring at very large distances or very long ago and certainly not locally.  It can be that what  you see isn't what you necessarily get.
\vskip 1in
\leftline{“Essay written for the Gravity Research Foundation 2020 Awards for Essays on Gravitation.”}
\leftline{\today}
\leftline{paranj@lps.umontreal.ca}
\end{abstract}
\maketitle
\section{Introduction}     
Causality is a fundamental principle that seems to concord with all our observations.  There is a clear understanding confirmed by experience that a given event can only affect those events that occur later than the causative event.   It is understood that this principle can be violated by certain space-times, if we admit matter that violates physically reasonable conditions.  For example, the G{\"o}del metric is a solution of the Einstein equations with a source corresponding to a negative cosmological constant and pressureless dust.   The G\"odel metric \cite{RevModPhys.21.447} admits closed time-like curves, {\ie}, it is possible to perform a circuit of this space-time always staying inside the locally defined future directed light-cone, and eventually return to the starting point.  Indeed, such a possibility is a gross violation of our notion of chronologicity, and has simply not been observed.  Notwithstanding, in general relativity it is easy to generate space-times which admit closed time-like curves.   Hawking \cite{PhysRevD.46.603} has even argued that closed time-like curves would be disallowed by a putative quantum gravitational effects, giving rise to his ``chronology protection conjecture''.   Closed time-like curves are certainly unphysical and perhaps quantum gravitational effects forbid them from entering our macroscopic, physical world.  However it would still be unnerving to see a system evolving backwards in time, even in the absence of closed time-like curves.   Such a possibility does exist, does not violate causality and of course requires the full complexity of general relativity and space-time curvature.  

\begin{figure}[ht]
\centerline{\includegraphics[scale=.5]{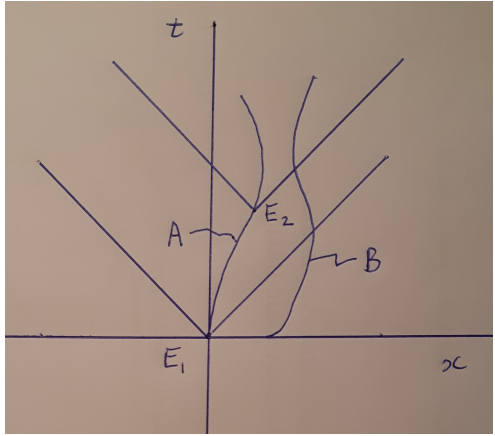}} \caption{\label{fig1} (color online)  Light-cones emanating from $E_1$ and from $E_2$ along the world line of $A$ and the world line of $B$ an arbitrary observer impinging on the two light-cones.}
\end{figure}

We can first dispense the possibility that without space-time curvature, there is no possibility of anti-chronological observation.  Consider two events $E_1$ and $E_2$, where $E_2$ is future time-like separated from $E_1$ in the coordinate system of  observer $A$.  
Now consider a second observer $B$ who is at the position $X_B$ at $t=0$ in the coordinates of $A$ (when event $E_1$ occurs according to $A$).  $B$ observes events $E_1$ and $E_2$ from the light that they  emit (assume everything is happening in a well lit room) as shown in Fig\eqref{fig1}.  Can $B$ see event $E_2$ before event $E_1$?  The answer is clearly no.
In fact, we can see from Fig.\eqref{fig1} that the world line of any observer, with arbitrarily accelerated motion, that starts outside the light-cone emanating from $E_1$ can never see $E_2$ before $E_1$.  This is because that observer's world line must first cross the light-cone from $E_1$ before it can cross the light-cone emanating from $E_2$.  

The underlying mathematical reason is that the ensemble of light cones emanating from the world line of  observer $A$, who travels from $E_1$ to $E_2$  gives a foliation of the full future directed light-cone emanating from $E_1$, specifically because the space-time is Minkowski space.  The world line of absolutely, any other time-like observer $B$, must intersect the leaves of the foliation sequentially and hence chronologically with respect to his or her proper time.  Thus the observer $B$ will see $E_1$ and $E_2$ in the same chronological order in which they occurred for observer $A$.  We will see that in general relativity this is not necessarily the case.
\section{Chronology in general relativity}
In general relativity, we can always have a null foliation of the local (compact) region of space-time, because of the equivalence principle.  Locally, space-time is simply homeomorphic to an open set in $\R^4$, but globally, anything can happen.  Null surfaces can intersect with each other or even with themselves.  Therefore even if you have a perfectly well defined, chronologically ordered set of events locally, how they are seen from far away is a matter of the intervening space-time curvature.   We will demonstrate this though a specific example, for technical details of the existing calculations that we will make use of, see \cite{Weinberg:1972kfs}. 

\begin{figure}[ht]
\centerline{\includegraphics[scale=.3]{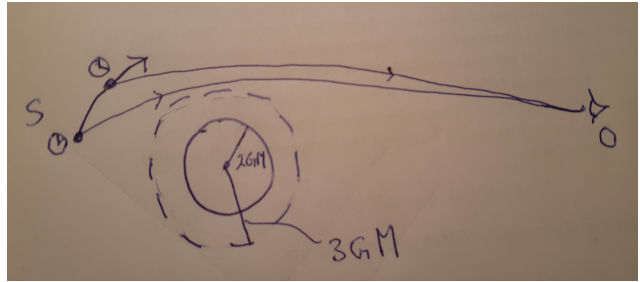}} \caption{\label{fig2} (color online)Light emitted by the system at an initial time and then somewhat later, and both rays arriving at the observer.}
\end{figure}

Consider a system $S$ and an observer $O$ in the presence of a black hole for convenience moving on circular trajectories.  The system could be a clock or something else that evolves in time, placed inside a spaceship that moves on a circle of radius $R_S$ however we permit the rocket to move at any speed not just the orbital velocity.  The observer is considered to be sufficiently far away, effectively at infinity.  We will do the analysis in Schwarzschild coordinates.  The black hole metric is given by (in units where $c=1$)
\beq
d\tau^2=\lb 1-\frac{2GM}{r}\rb dt^2-\lb 1-\frac{2GM}{r}\rb^{-1}dr^2-r^2d\theta^2-r^2\sin^2\theta d\phi^2
\eeq
and we assume the motion occurs in the plane with $\theta=\pi/2$.
The system of mass $M_S$ is at radius $R_S$ while the observer of mass $M_O$ is at radius $R_O$, with $R_O\gg R_S$.  The black hole has mass $M$,  the  observer is far away from the black hole so $R_O\gg GM$ and additionally $M\gg M_S, M_O$.  Being far away from the black hole,  the observer's proper time and the evolution of coordinate time are arbitrarily close
\beq
d\tau_O\approx d t.
\eeq
The system on the other hand will be allowed to be quite close to the black hole.  
A light ray arriving from the system to the observer will suffer the Shapiro time delay, interestingly, this delay can be arbitrarily large.  Consider the situation as pictured in Fig.\eqref{fig2}.
The problem has been well studied,
using a dimensionless variable $\rho$ with $r=GM\rho$, $r_0=GM\rho_0$ (the point of closest approach), $R_S=GM\rho_S$ and $R_O=GM\rho_O$, the lapse of coordinate time for the light ray to pass from the system to the point of closest approach and then on to the observer is
\beq
t_{\rm Shapiro}(\rho_S,\rho_0)=GM\left(\int_{\rho_0}^{\rho_S}+\int_{\rho_0}^{\rho_O}\right)d\rho\frac{\rho^{5/2}}{\rho-2}\lb\frac{\rho_0-2}{\rho^3\lb\rho_0-2\rb-\rho_0^3\lb\rho-2\rb}\rb^{1/2}.\label{tshap}
\eeq
We note $\rho=2$ is the event horizon and $\rho=3$ is the photosphere.   The key function in the integrand is the denominator inside the square root $D(\rho)=\rho^3\lb\rho_0-2\rb-\rho_0^3\lb\rho-2\rb$ which obviously vanishes at $\rho=\rho_0$ and the other roots are at  $\rho=\rho_\pm= \frac{\rho_0}{2}\lb-1\pm\sqrt\frac{\rho_0+6}{\rho_0-2}\rb$, one of which is obviously negative.  For $\rho_0>3$ the largest root is $\rho_0$ while $\rho_+=\frac{\rho_0}{2}\lb-1+\sqrt\frac{\rho_0+6}{\rho_0-2}\rb\le\rho_0$ with equality attained for $\rho_0=3$.  For $\rho_0\in\lb2,3\rb$, $\rho_+$ and $\rho_0$ exchange their order and for the $\rho\in (\rho_0,\rho_+)$ the sign of $D(\rho)$ becomes negative and there be dragons!  We will not entertain this possibility here.

For $\rho>3$ the single zero in the denominator at $\rho=\rho_0$ manifests as an integrable singularity and poses no problem for the integration.  For $\rho_0=3$ the single root becomes a double root and the denominator becomes $D(\rho)=\rho^3-27\lb\rho-2\rb=\lb\rho-3\rb^2\lb\rho+6\rb$.  Thus taking account of the square root, as $\rho_0\to 3^+$   a logarithmic divergence appears in the integration and the Shapiro time delay can be arbitrarily large.
 
Consider the system first at a position so that the observer would see it only if the light ray just grazes the photosphere, the Shapiro time delay is arbitrarily large.  Then as the system continues in its circular trajectory, the Shapiro time delay would become smaller and smaller till effectively it becomes negligible.  It is possible that the observer could at some point see the system at its later position before the observer sees it at its earlier position.  There will be critical photon trajectory 
\section{Critical photon trajectory}
The critical photon trajectory will correspond to the position at which light signals emitted from the system at coordinate time $t$ and $t+dt$ arrive at the observer simultaneously.  From the critical trajectory, two images of the system will emanate, one evolving forward in time and the other evolving backwards in time.  

The time on a clock in the system will show its proper time $\tau_S$ at coordinate time $t$.  Its arrival time at the observer will be $t+t_{{\rm Shapiro}}(\rho_S,\rho_0(t))$.  At $t+dt$ coordinate time, the clock in the system will show its proper time advanced to $\tau_S+d\tau_S$, and a light ray emitted by the system then, will be seen by the observer at $t+dt +t_{{\rm Shapiro}}(\rho_S,\rho_0(t+dt))$. 
The critical photon trajectory will the one for which these two light rays arrive simultaneously at the observer
\beq
t+dt +t_{{\rm Shapiro}}(\rho_S,\rho_0(t+dt))=t+t_{{\rm Shapiro}}(\rho_S,\rho_0(t))
\eeq
which gives
\beq
\frac{dt_{{\rm Shapiro}}(\rho_S,\rho_0(t))}{dt}=-1.
\eeq
We observe that $t_{{\rm Shapiro}}$ contains a term linear in $R_O$ and $R_S$, a logarithmically divergent term as $\rho_0\to\rho_+$ for $\rho_0\to 3$ and additional regular terms.  Writing
\beq
t_{{\rm Shapiro}}=t_{\rm linear}+t_{\rm log}+t_{\rm regular}
\eeq
we have
\beq
t_{\rm linear}\approx\sqrt{\rho_S^2+\rho_0^2-2\rho_0\rho_S\cos\phi_S}+\sqrt{\rho_O^2+\rho_0^2-2\rho_0\rho_O\cos\phi_O}
\eeq
where $\phi_S$ and $\phi_O$ are the angular coordinates of the system and the observer respectively.  We use the approximation that the spacetime is essentially flat except for very near the black hole.  $\phi_S=\omega t$ while $\phi_O$ is assumed to be essentially constant.  We will not require the explicit form of $t_{\rm linear}$, only that it is a decreasing function of $t$.

To find the dependence of $\rho_0(t)$ (on $t$) is difficult however, it is possible to use the deflection angle (see \cite{Weinberg:1972kfs}), $\Delta\Phi (\rho_0)$ 
\beq
\Delta\Phi (\rho_0)= \rho_0^{3/2}\left(\int_{\rho_0}^{\rho_S}+\int_{\rho_0}^{\rho_O}\right)d\rho\frac{1}{\rho^{1/2}}\lb\frac{1}{\rho^3\lb\rho_0-2\rb-\rho_0^3\lb\rho-2\rb}\rb^{1/2}.\label{deltaphi}
\eeq
We imagine the system is moving while the observer is essentially motionless which occurs for $\rho_O\gg \rho_S$.  The deflection angle and $\rho_0$ vary as the system moves.   Using the implicit inverse relation $\rho_0(\Delta\Phi)$ we have 
\beq
\frac{dt_{{\rm Shapiro}}(\rho_S,\rho_0(t))}{dt}=\frac{dt_{{\rm Shapiro}}(\rho_S,\rho_0)}{d\rho_0}\frac{d\rho_0(\Delta\Phi )}{d\Delta\Phi }\frac{d\Delta\Phi(t)}{dt}=\frac{\frac{dt_{{\rm Shapiro}}(\rho_S,\rho_0)}{d\rho_0}}{\frac{d\Delta\Phi (\rho_0)}{d\rho_0}}\frac{d\Delta\Phi(t)}{dt}\label{20}
\eeq
where simply $\frac{d\Delta\Phi(t)}{dt}=-\omega$ (note that $\Delta\Phi$ decreases as the system moves).

We will use Eqn.\eqref{20} to calculate $\frac{dt_{{\rm log}}(\rho_S,\rho_0(t))}{dt}$.  Both $t_{{\rm log}}(\rho_S,\rho_0)$ and $\Delta\Phi (\rho_0)$ diverge logarithmically as $\rho_0\to 3$.  Writing the denominator appearing in either one,  and making a change of variables $\rho=\rho_0 s$, we have
\beq
\lb\frac{1}{\rho^3\lb\rho_0-2\rb-\rho_0^3\lb\rho-2\rb}\rb^{1/2}=\lb\frac{1}{\rho_0^3(\rho_0-2)(s-1)(s-\rho_+/\rho_0)}\rb^{1/2}\lb\frac{1}{s-\rho_-/\rho_0}\rb^{1/2}
\eeq
we note that $\rho_+\to 3$ when $\rho_0\to 3$ yielding a logarithmic divergence from the first term, while the remaining factor is analytic as $s\to 1$.  Thus both integrals Eqn.\eqref{tshap} and Eqn.\eqref{deltaphi}  have the form
\beq
\int_1^{\frac{\rho_{S,O}}{\rho_0}}ds\lb\frac{1}{(s-1)(s-\rho_+/\rho_0)}\rb^{1/2}f(\rho_0, s)
\eeq  
where $f(\rho_0, s)$ is analytic at $s=1$ for all values of $\rho_0\ge 3$.  Thus the contribution to the integral, that which diverges logarithmically, will be obtained by replacing $f(\rho_0, s)\to f(\rho_0, 1)$.  Subsequent terms in the Taylor expansion of $f(\rho_0, s)$ will only contribute finite, negligible terms if $\rho_0$ is near 3.  When $f(\rho_0, s)\to f(\rho_0, 1)$ the remaining integral can be done analytically
\beq
\int ds\lb\frac{1}{(s-1)(s-\rho_+/\rho_0)}\rb^{1/2}= 2\ln\lb \sqrt{{s-1}}+\sqrt{{s-\rho_+/\rho_0}}\rb + {\rm constant} .
\eeq
Evaluating at the limits, we obtain the logarithmically diverging contribution
\beq
\int_1^{\frac{\rho_{S,O}}{\rho_0}}ds\lb\frac{1}{(s-1)(s-\rho_+/\rho_0)}\rb^{1/2}f(\rho_0, s)=-f(\rho_0, 1)\ln\lb\rho_0-\rho_+\rb + {\rm regular\,\,\, terms}.
\eeq
Then the derivatives required in Eqn.\eqref{20} will only give the dominant contribution when the logarithm is differentiated and the ensuing derivatives cancel, giving the condition for the critical light ray
\beq
\frac{dt_{{\rm Shapiro}}(\rho_S,\rho_0(t))}{dt}=\frac{d\lb t_{\rm linear}+t_{\rm regular}\rb}{dt}-\frac{f_{t_{\rm log}}(\rho_0,1)}{f_{\Delta\Phi}(\rho_0,1)}\omega\approx \frac{d\lb t_{\rm linear}\rb}{dt}-\frac{f_{t_{\rm log}}(\rho_0,1)}{f_{\Delta\Phi}(\rho_0,1)}\omega=-1.\label{28}
\eeq
It is straightforward to determine  $f_{t_{\rm log}}(\rho_0,1)$ and $f_{\Delta\Phi}(\rho_0,1)$, for each of the integrations from $\rho_0$ to the system and from $\rho_0$ to the observer,  we find 
\beq
f_{t_{\rm log}}(\rho_0,1)=   2\frac{GM\rho_0^{5/2}}{(\rho_0-2)(\rho_0-\rho_-)^{1/2}}           \quad\quad f_{\Delta\Phi}(\rho_0,1)=\frac{2\rho_0}{(\rho_0-2)^{1/2}(\rho_0-\rho_-)^{1/2}}\label{29}
\eeq
Replacing into Eqn.\eqref{28} and changing the overall sign we find (we note that $ \frac{d\lb t_{\rm linear}\rb}{dt}$ is negative)
\bea
\left| \frac{d\lb t_{\rm linear}\rb}{dt}\right|+2\frac{GM\rho_0^{5/2}}{(\rho_0-2)(\rho_0-\rho_-)^{1/2}} \frac{(\rho_0-2)^{1/2}(\rho_0-\rho_-)^{1/2}}{2\rho_0}\omega=\nonumber\\
=\left| \frac{d\lb t_{\rm linear}\rb}{dt}\right|+(GM\omega)\frac{\rho_0^{3/2}}{ (\rho_0-2)^{1/2}}=1.\label{criticalray}
\eea
We will show that the second term can vary from $0\to 1$ for $\omega: 0\to\omega_{\rm max}$ (defined when the system moves at the speed of light) and $\rho_S\to\rho_0$.   The first term is positive and also vanishes for $\omega=0$.  Hence there will always be a physical solution for $\omega<\omega_{\rm max}$ for which the Eqn.\eqref{criticalray} is satisfied, {\ie.} a critical photon trajectory.

For movement on the circular trajectory at radius $\rho_S$ at $\theta=\frac{\pi}{2}$ we have
\beq
d\tau^2=\lb 1-\frac{2}{\rho_S}-(GM\rho_S)^2\omega^2\rb dt^2
\eeq
and for this motion to be time-like we must have
\beq
(GM\omega)^2\le\frac{1}{\rho_S^2}\lb1-\frac{2}{\rho_S}\rb = \frac{(\rho_S-2)}{\rho_S^{3}},
\eeq
with the inequality saturated for 
\beq
\omega_{\rm max}=\frac{1}{GM}\frac{(\rho_S-2)^{1/2}}{\rho_S^{3/2}}
\eeq
when the system is moving at the speed of light.  As $\rho_0\le\rho_S$ and $\frac{(\rho_S-2)^{1/2}}{\rho_S^{3/2}}$ is a strictly decreasing function of 
$\rho_S$, we have for the contribution of the second term in Eqn.\eqref{criticalray}
\beq
(GM\omega)\frac{\rho_0^{3/2}}{ (\rho_0-2)^{1/2}}\le \frac{(\rho_S-2)^{1/2}}{\rho_S^{3/2}}\frac{\rho_0^{3/2}}{ (\rho_0-2)^{1/2}}\le\frac{(\rho_0-2)^{1/2}}{\rho_0^{3/2}}\frac{\rho_0^{3/2}}{ (\rho_0-2)^{1/2}}= 1.
\eeq
Thus as $GM\omega$ varies from $0\to \frac{(\rho_0-2)^{1/2}}{\rho_0^{3/2}}$ the second term in Eqn.\eqref{criticalray}
varies from $0 \to 1$.  Hence for any value of $\left| \frac{d\lb t_{\rm linear}\rb}{dt}\right|$, (which behaves as $\sim {\rho_0}\omega$), there is always a solution for $\omega$ corresponding to time-like motion (slower than the speed of light), for which Eqn.\eqref{criticalray} is satisfied.  Therefore the critical photon trajectory always exists.  

\section{Discussion}

Let us examine what this implies for the observations.  Suppose the observer first sees the system exactly at the critical photon trajectory.  Continuously and smoothly afterwards, the observer will see two images of the system, one continuing onward in its circular trajectory, while the other moving backwards along the circular trajectory.  The system, on the backwards trajectory will be seen by the observer to be evolving backwards in time, {\it ie. anti-chronologically}, while on the forward trajectory it will be seen by the observer to be evolving forwards in time,  {\it ie. chronologically}. 

This observation is not the same as seeing say two images of an evolving astrophysical system such as a pulsar, one image arriving even hours after the first.  Here we are demonstrating the possibility of a differentiable continuum of a sequence of images that will appear as if the system evolved backwards in time.  

Chronologicity is not a globally defined concept in general relativity.  We have shown that we can see a clock running backwards in time.  This is not just a movie that is being played in reverse, one can actually see a later event (later in the system's reference frame) before we see an earlier event(earlier in the system's reference frame).   There is no violation of causality as we are not in the local neighbourhood of the system.  Space-time creates an optical illusion.  

This may not be so nerve wracking, however, imagine the processes that we can see.  For example we know that time reversal invariance is violated at a fundamental level in the weak interactions of particle physics \cite{Quinn:2009zza}.  This means that there are processes that do happen going forward in time, however, the time reversed process is forbidden.  With the situation described above, we can {\it see} such forbidden processes to occur.  As another example, even at the classical level, the collapse of a neutron star into a black hole is a classical process which does happen and is observed.   The observation of its time reversal is anathema, not realistically credible.  However one might {\it see} such a thing, a black hole will shrink and give out matter until the original supernova that caused it to be created, appearing to evolve backwards in time!  

Although there may be no realistic aspiration that we would actually {\it see} such things, it is important to know that such strange effects are in principle observable.  If all observations are being done of systems that are located far away, it might be a good idea to reflect upon whether what you see is really what is happening.

\section{Acknowledgments} We thank NSERC of Canada for financial support.  We thank Naresh Dadhich, Sunil Mukhi, Cliff Burgess, Emil Mottola for informative discussions.

\bibliographystyle{apsrev}
\bibliography{causality-locality}

\end{document}